# The Mass-to-Light Ratios of the Draco and Ursa Minor Dwarf Spheroidal Galaxies. II. The Binary Population and its Effect on the Measured Velocity Dispersions of Dwarf Spheroidals[1]

Edward W. Olszewski[2]

Steward Observatory, University of Arizona, Tucson, AZ 85721

Email address: eolszewski@as.arizona.edu

Carlton Pryor[2]

Dept. of Physics and Astronomy, Rutgers, the State University of New Jersey,
P.O. Box 849, Piscataway, NJ 08855–0849

Email address: pryor@physics.rutgers.edu

and

Taft E. Armandroff

National Optical Astronomy Observatories, Kitt Peak National Observatory,
P.O. Box 26732, Tucson, AZ 85726

Email address: tarmandroff@noao.edu

## ABSTRACT

We use a large set of radial velocities in the Ursa Minor and Draco dwarf spheroidal galaxies to search for binary stars and to infer the binary frequency. Of the 118 stars in our sample with multiple observations, six are velocity variables with $\chi^2$ probabilities below 0.001. We use Monte Carlo simulations that mimic our observations to determine the efficiency with which our observations find binary stars. Our best, though significantly uncertain, estimate of the binary frequency for stars near the turnoff in Draco and UMi is 0.2–0.3 per decade of period in the vicinity of periods of one year, which is 3–5× that found for the solar neighborhood. This frequency is high enough that binary stars might significantly affect the measured velocity dispersions of some dwarf spheroidal galaxies according to some previous numerical experiments.

However, in the course of performing our own experiments, we discovered that this previous work had inadvertently overestimated binary orbital velocities. Our first set of simulations of the effects of binaries is based on the observed scatter in the individual velocity measurements for the multiply-observed Draco and Ursa Minor stars. This scatter is small compared to measured velocity dispersions and, so, the effect of binaries on the dispersions is slight. This result is supported by our second set of experiments, which are based on a model binary population normalized by the observed binary frequency in Draco and Ursa Minor. We conclude that binary stars have had no significant effect on the measured velocity dispersion and inferred mass-to-light ratio of any dwarf spheroidal galaxy.

*Subject headings:* Dwarf Spheroidal Galaxies, Stellar Systems (Kinematics, Dynamics), Binary Stars, Dark Matter, Radial Velocities, Carbon Stars

## 1.  Introduction

The measured velocity dispersions of dwarf spheroidal galaxies are larger than predictions based on their central surface brightness, simple dynamical models, and the assumption that the mass-to-light ratio should be like that of globular clusters. This has led to the suggestion that dwarf spheroidal galaxies contain significant amounts of dark matter, making them the smallest stellar systems in which dark matter is found (for recent reviews see Mateo 1994 and Pryor 1994). The two dwarf spheroidal galaxies that are the most extreme in this regard are Draco and Ursa Minor (UMi), with velocity dispersions of about 10 km s$^{-1}$ and V-band mass-to-light ratios from King model fits of 50–90 (Armandroff *et al.* 1995; Olszewski *et al.* 1995). For reference, globular clusters have V-band mass-to-light ratios of 1–3 (*e.g.,* Pryor & Meylan 1993).

A number of authors have suggested that the orbital velocities of undetected binary stars may artificially inflate the observed velocity dispersions of dwarf spheroidal galaxies. The dwarf spheroidals with only single-epoch velocity measurements and smaller velocity dispersions are the systems most vulnerable to the effects of binaries. The binary frequency of Population II stars is becoming better known through work on globular clusters (Hut *et al.* 1992) and field stars (Latham *et al.* 1992, 1988) and appears to be, at most, only slightly smaller than the high frequency for the Population I field. In order to investigate the effects of undetected binary stars on the measured velocity dispersions of the dwarf spheroidals, a number of authors have performed Monte Carlo simulations of velocities measured at random times for a sample of stars with a given velocity dispersion, binary frequency, period distribution, ellipticity distribution, and secondary mass distribution (Mateo *et al.* 1993; Suntzeff *et al.* 1993; Vogt *et al.* 1995). These simulations have shown that: 1) The velocity dispersion calculated using the standard deviation is much more vulnerable to the effects of binaries than that calculated from the biweight, a robust estimator (Beers *et al.* 1990); 2) Very high binary frequencies or binaries with periods restricted to certain ranges *may* be able to partially explain the large dispersions in the dwarfs with smaller velocity dispersions (*i.e.*, $\sigma \approx 6$ km s$^{-1}$) and only single-epoch velocity measurements; and 3) Multi-epoch velocity measurements are desirable since they allow binaries with large velocity amplitudes to be detected and eliminated. The choices of the period distribution, ellipticity distribution, and secondary mass distribution in these simulations are guided by observed results for field binaries. However, the frequency of binaries in dwarf spheroidals is a free parameter that exerts considerable influence over the results.

During the past 12 years, we have measured precise radial velocities for stars in the Draco and Ursa Minor dwarf spheroidal galaxies using the Multiple Mirror Telescope (MMT) with the echelle spectrograph (Olszewski *et al.* 1995) and the KPNO 4-m telescope with the Hydra multi-fiber spectrometer (Armandroff *et al.* 1995). In addition, Hargreaves *et al.* (1994b) have measured velocities in UMi. Combining these three studies, there are 548 velocities for 185 members of these two galaxies. These multi-epoch observations permit the removal of radial-velocity variables from the sample, thus minimizing the effects of binary stars on the measurement of the velocity dispersion. In addition, this set of velocities allows us to investigate the frequency of binary stars in dwarf spheroidal galaxies for the first time. In this paper, we apply the techniques employed to estimate the binary frequency among field stars (Duquennoy & Mayor 1991) and globular cluster stars (Hut *et al.* 1992) to the combined set of velocities for Draco and UMi. This results in a binary frequency for these galaxies, which is of intrinsic interest and can be compared to the Pop I and other Pop II samples. We then use this empirically-determined dwarf-spheroidal binary frequency to investigate the effects of the binary population on velocity dispersions derived from single-epoch velocities.

We will conclude that binaries are not responsible for the large measured mass-to-light ratios in dwarf spheroidals. We will discuss extensively the kinematics, masses, and mass-to-light ratios of UMi and Draco in the third paper of this series (Pryor *et al.* 1996).

The paper is organized as follows. The next section briefly describes our database of velocities, the criterion for velocity variability, and our binary candidates. Section 3 presents our analysis of the binary frequency in Draco and UMi. Our numerical experiments that evaluate the impact of binaries on the measured velocity dispersions, both empirical and model-based, are discussed in Sec. 4. We summarize our conclusions in Sec. 5.

## 2.  The Data and the Binary Candidates

The data are radial velocities of stars in the Draco and Ursa Minor dwarf spheroidal galaxies obtained using the MMT echelle (Olszewski *et al.* 1995, hereafter OAH95), the Kitt Peak 4m with the Hydra multi-fiber positioner and bench spectrograph (Armandroff *et al.* 1995, hereafter Paper I), and the William Herschel Telescope with ISIS (Hargreaves *et al.* 1994b, hereafter H94). Paper I and OAH95 describe the data, the uncertainties in the individual velocities, and the comparisons between the different sets of data in great detail. We will repeat some of the essential points here.

The MMT data consist of 112 velocities for 42 stars, with observations between April 1982 and Sept. 1990. A single order of the MMT echelle was observed, covering approximately $\lambda\lambda$ 5160–5213 Å for the K giants or $\lambda\lambda$ 5600–5670 Å for the Carbon (C) stars. The velocity resolution is 11 km s$^{-1}$ and the typical uncertainty in a single measurement is 1.7 km s$^{-1}$. H94 observed at the calcium triplet, approximately $\lambda\lambda$ 8300–8750 Å, with a spectral resolution of 25 km s$^{-1}$. They obtained 63 velocities with Tonry-Davis (1979) $R > 7.5$ for 34 UMi stars in 1991 and 1992. The uncertainty which we determined from H94's repeat measurements in Paper I is 2.3 km s$^{-1}$. However, we also argued in Paper I that there is an additional additive uncertainty of 3.1 km s$^{-1}$, giving a total uncertainty of 3.9 km s$^{-1}$ (see below). The Hydra data in Paper I is from the years 1992, 1993, and 1994, yielding 373 velocities for 185 probable members. The spectral coverage was approximately $\lambda\lambda$ 4720–5460 Å, the resolution 70 km s$^{-1}$, and the median velocity uncertainty 4.0 km s$^{-1}$.

In order to use these three datasets to search for binary stars, they must be placed on a common velocity zero-point and the uncertainties must be correctly estimated. Paper I again contains an exhaustive discussion of these issues. Based on the stars in common, in Paper I we applied small velocity zero-point shifts to the Hydra and H94 data to put them on the same system as the OAH95 data, which have the highest resolution. The velocity uncertainties for each individual dataset were then calculated by using the repeat measurements within that dataset. Our basic tool for this work was the $\chi^2$ of the scatter of the velocities for each star around its weighted mean velocity and the probability of seeing a value of $\chi^2$ larger than was observed. If the uncertainties are correct, the $\chi^2$ probabilities will be uniformly distributed between 0.0 and 1.0. In order to prevent the presence of velocity variables from artificially inflating the measurement errors, we require a uniform distribution only for those stars with $\chi^2$ probabilities above 5% (a method first used in Duquennoy & Mayor 1991). Departures from a uniform distribution are only seen below probabilities of about 1% (see Fig. 3 of Paper I), so the 5% limit is conservative.

We showed in Paper I that the internal uncertainties of the MMT and Hydra velocities are realistic and this conclusion was further supported by comparing the two sets of data. Comparing the H94 velocities with those from the MMT and Hydra showed that the H94 data seem to suffer from a rather large additive error that was the same for all H94 measurements of a given star, but varied randomly from star to star. The mean systematic error was included as an additional uncertainty added in quadrature to the uncertainties of the H94 velocities. Section 5.2 of Paper I contains an extensive discussion of this problem.

The work outlined above resulted in the individual and mean velocities for the 185 likely members of UMi and Draco found in Tables 2 and 3 of Paper I. These stars all have velocities within about 50 km s$^{-1}$ of the mean radial velocity of each galaxy. Figures 11 and 12 of Paper I and the discussion accompanying them show that this choice of stars is reasonable. The stars with velocities farthest from the mean of their galaxy either have only a single measurement or have two measurements that agree well, so the inclusion or exclusion of these stars from the sample has no significant effect on the conclusions of this paper.

Though these data are the most extensive set of repeated measurements assembled for dwarf spheroidal galaxies, they are still much too sparse to even attempt to deduce binary orbital properties for any but perhaps the few most heavily observed stars. We are thus forced to the cruder technique of identifying likely velocity variables and using simulations of the observations to determine what these data tell us about the population of binary stars in the galaxies (*e.g.*, Pryor *et al.* 1989, Hut *et al.* 1992). The wide range of velocity uncertainties in the data, from 1 to 10 km s$^{-1}$, make a "discovery criterion" for velocity variables based on the velocity range less useful than in the previous similar work on globular clusters. In this paper we use the probability of the $\chi^2$ of the scatter around the weighted mean velocity as the basis of our criterion, adopting as variables those stars with probabilities below 0.001.

We will demonstrate in the next section that all of the giants in our sample have radii that are too large to fit in binary systems with periods shorter than about 90 days. Thus the errors in the approximate Julian dates adopted for the Hydra and H94 velocities in Paper I, which could be as large as two days, are of no consequence for the detection of velocity variability. This restriction on the periods also means that repeated measurements within an H94 or Hydra run should measure the same velocity and so reduce the sensitivity for detecting binaries by decreasing the total $\chi^2$. We thus replaced the velocities in our sample that were separated by less than 7.0 days with their weighted mean. The Julian date adopted for the mean velocity was the average Julian date weighted by the velocity uncertainties. This also resulted in combining a few MMT velocities taken on adjacent nights. The result was a sample of 188 velocities for 59 stars with repeated measurements in Draco and 231 velocities for 59 stars with repeated measurements in UMi.

In order to derive the radii of the stars in our sample, we need their luminosities and effective temperatures. Photometry is also important in assessing the reality of our binary candidates. Measurements of the radial velocities of stars near the tip of the giant branch in globular clusters have shown that the velocities of such stars can vary because of atmospheric motions (Gunn & Griffin 1979; see the discussion in Hut *et al.* 1992). We draw this photometry from the published work of Cudworth *et al.* (1986) for UMi and that of Stetson (1979) for Draco. Additional values for Draco come from unpublished photometry by Stetson and McClure; a very few magnitudes come from Baade & Swope (1961) and Aaronson *et al.* (1983). The remainder comes from M. Irwin's APM photometry (see Paper I), transformed to V.

Figure 1 shows the probability that a star with a constant velocity would exceed the observed $\chi^2$ value by chance *vs.* the absolute visual magnitude for the 118 stars in the combined Draco and UMi sample. A similar plot is shown for the MMT data alone in OAH95. We adopt apparent distance moduli of 19.6 for Draco and 19.3 for UMi and discuss the source of these values in the next section. In these metal-poor systems, the tip of the giant branch should be at about $M_V = -2.6$ to $-2.7$ (Bergbusch & VandenBerg 1992, Rees 1992, Stetson & Harris 1988). There are six stars with probabilities below 0.001: stars 249 and C1 in Draco and COS60 (M), COS215 (K), N33, and VA335 in UMi; one star with a probability between 0.001 and 0.01: star JI8 in UMi; and six with probabilities between 0.01 and 0.05: stars XI–2, 24, 473, 3237 (C3), and 427 in Draco and JI12 in UMi. The expected numbers of stars in these probability ranges for a sample of 118 stars is 0.1, 1.1, and 4.7. Thus only for probabilities below 0.001 is there a clear excess of stars indicating the presence of truly variable velocities. Table 1 contains the velocity data for the six stars with probabilities below that limit. The data are extracted from Tables 2 and 3 of Paper I, with velocities separated in time by less than 7.0 days combined as discussed above.

The six stars with variable velocities do cluster near the giant branch tip in Fig. 1, which would suggest that at least some of their variability is due to atmospheric motions. Pryor *et al.* (1988) find that the largest atmospheric "jitter" of 4–8 km s$^{-1}$ occurs in stars within about 0.5 mag of the giant branch tip in globular clusters. However, only Dra 249 and UMi N33 have velocity ranges that are close to this size, 10.3 km s$^{-1}$ and 7.9 km s$^{-1}$, respectively. We can also be reasonably sure that the other four stars are binaries from other evidence. Both Dra C1 (Aaronson *et al.* 1982) and UMi M show strong emission lines in their spectra and C1 is a C star. UMi K and VA335 are also both C stars, likely

CH stars (Aaronson & Mould 1985). McClure (1984) and McClure & Woodsworth (1990) have shown that similar stars in the Milky Way are almost always binaries. Mayor *et al.* (1995) have recently found that the CH stars in $\omega$ Cen do not have a high binary frequency. These $\omega$ Cen CH stars do not show exceptionally large velocity jitter either, however, while the velocity ranges of UMi K and VA335 are large. The $\omega$ Cen result may be due to the disruption of binaries in the dense globular cluster environment (Mayor *et al.* 1995). The most problematic of the six velocity variables, Dra 249 and UMi N33, have absolute visual magnitudes of about −2.3, placing them 0.3–0.4 mag below the red giant branch tip. Given their large velocity ranges, we believe that the balance of the evidence favors both of these stars being binary candidates. The exclusion of these two stars will only affect the derived binary fraction (see Table 2) and the simulation in Sec. 4 that is based on the binary fraction.

It would be useful to know whether these stars near the tip are photometric variables, but the evidence is unfortunately quite sparse. Neither Baade & Swope (1961) nor van Agt (1967) found any red irregular or long-period variables in Draco or Ursa Minor. H. Harris (private communication) and collaborators have begun a new search for variables in Draco, and to date report one red variable (see the discussion below).

One reason for the apparent concentration of velocity variables near the tip may be that the brightest stars have the largest number of and the most precise velocities. These data are the most efficient at identifying velocity variables with a $\chi^2$ probability criterion. For example, the stars with absolute visual magnitudes brighter than −2.0 have an average uncertainty in the individual measurements of 2.5 km s$^{-1}$ and an average of 4.5 measurements per star. The averages for the fainter stars are 4.8 km s$^{-1}$ and 2.6. Simulations of the efficiency with which the velocity data can discover binaries, similar to those described in the next section, show that the efficiency is about 3× larger for the stars brighter than an absolute magnitude of −2.0 than for those fainter.

A final approach to exploring the question of velocity variability due to atmospheric motions among the Draco and UMi stars is to examine the distribution of stars in Fig. 1. Pryor *et al.* (1988) found that stars in the globular cluster M3 with absolute visual magnitudes brighter than −2.0 preferentially had $\chi^2$ probabilities below 0.5, while there were roughly equal numbers of stars above and below this value at fainter magnitudes. They attributed this effect to velocity "jitter" among the brighter stars. As OAH95 found for the MMT Draco and UMi data alone, we see roughly equal numbers of stars in all 4 quadrants of Fig. 1. For stars brighter than −2.0, we see 31 (26) stars with probabilities less than 0.5, and 27 (26) stars with probabilities greater than 0.5 (the numbers in parenthesis reflect the elimination of C stars and the stars with emission lines). Similarly, for the fainter sample, stars with probabilities less than 0.5 number 37 (35), while the more probable stars number 23 (22). Either velocity jitter is unimportant in our sample, or it is being masked by the uncertainties in the observations. In either case, the stars Dra 249 and UMi N33 are more likely to be actual binaries than stars with atmospheric jitter.

Although our six stars with $\chi^2$ probabilities less than 0.001 have only 4–7 velocities each, some interesting limits can be placed on the orbital periods of the stars. Figure 2 shows the velocities of all six stars as a function of time. We attempted to fit sinusoids to the velocities of three stars: UMi M and Draco C1, with 7 velocities each; and UMi N33 with 6 velocities. They were observed over intervals of 3714, 4051, and 2538 days, respectively. UMi M could not be fit; for any period and amplitude of motion the lowest $\chi^2$ was of order 30. We suspect that it is in an elliptical orbit, as a change of 30 km s$^{-1}$ in 89 days with much smaller changes from then on would hint. The star vZ 164 in M3 shows a qualitatively similar behavior (see Fig. 2 of Hut *et al.* 1992). Many more velocities will be needed to understand the orbit of this likely binary.

The presence of strong emission lines suggests that Draco C1 is likely to be filling its Roche lobe, which constrains the range of orbital periods and thus the fitted sinusoids. For a star at the tip of the red giant branch, with a radius of 0.40 AU, the critical separation for mass transfer given by the approximation of Eggleton (1983) is 1.06 AU (assuming a mass ratio of 1.0, but the function varies slowly with ratio). For a primary mass, $M_1$, of 0.8 M$_\odot$ and a secondary mass, $M_2$, of 0.6 M$_\odot$, the period is 335 days and the velocity amplitude for a circular orbit is 14.8 km s$^{-1}$. If C1 is slightly below the RGB tip with a radius of 0.3 AU, $M_1$=0.7 M$_\odot$, and $M_2$=0.5 M$_\odot$, then the critical separation for mass transfer is 0.74 AU. That size yields a period of 210 days and a velocity amplitude for a circular orbit of 16.0 km s$^{-1}$. Our velocities allow three different periods for Draco C1 equal to or longer than this minimum: $\sim 180^d$, $\sim 375^d$, and $\sim 570^d$. The velocity curve for $375.5^d$ is the sinusoid in Panel 5 of Fig. 2. Draco C1 is also a photometric variable with amplitude $\geq 0.25$ mag (H. Harris, private communication). Note that our velocities come from absorption lines, mostly from the C$_2$ bandheads, and thus we are probably measuring orbital motion. This C star is not the luminous type of C star found in the dwarf spheroidals with young populations or in the disk of the Milky Way.

UMi N33 has periods of shorter than $200^d$ ruled out by its size. If it is a binary, the only reasonable periods in the interval 150–600$^d$ are $\sim 200^d$ and $\sim 220^d$. The velocity curve for 216.7 days is the sinusoid in Panel 6 of Fig. 2. Clearly, additional velocities measured over the next decade are needed to learn more about these three stars.

### 3. Binary Star Detection Efficiencies and the Binary Frequency

The six binary star candidates with $\chi^2$ probabilities below 0.001 imply a "binary discovery fraction" of 0.051. The 95% confidence interval for this fraction is (0.024, 0.096). This interval results from asking how small (large) the fraction could be and still have a 2.5% chance of seeing six or more (six or less) stars with probabilities below 0.001 in a sample of 118 stars. If Dra 249 and UMi N33 are excluded as binary candidates, the discovery fraction and its confidence interval are 0.034 and (0.0094, 0.084).

One might be tempted to argue that the three C stars with $\chi^2$ probabilities below 0.001 should be removed from this sample, lowering the discovery fraction and the eventual binary population inferred by a factor of two. However, of the nine C stars known in Draco and UMi (Paper I), we have counted as binary candidates only those three which met our velocity variability criterion. The C stars found in Draco and Ursa Minor are most likely to be CH stars (Aaronson & Mould 1985) and are not the product of intermediate-mass single-star evolution. UMi and Dra have stellar populations that are quite old (Olszewski & Aaronson 1985; Stetson *et al.* 1985; Carney & Seitzer 1986) and rather similar to the populations in globular clusters and in the Galactic halo. While UMi and Dra are the least luminous dwarf spheroidals (e.g., Mateo 1994), they have luminosities comparable to those of the most luminous globulars. Indeed, $\omega$ Cen, M22, M55, and M2 contain CH stars. It seems clear that no matter what mechanism makes CH stars, those that are binaries are simply part of the normal mix of binaries in these populations. Removing the C stars from our sample is therefore inappropriate when comparing the true number of binaries present in different samples. As we discuss in Sec. 4, when we test for the effect of binaries on the measured velocity dispersions of dSphs, it is appropriate to remove the C stars because, in general, C stars are now avoided when measuring velocities for deriving dispersions.

In order to go from the discovery fraction to the fraction of stars in Draco and UMi that are actually binaries, the "binary frequency", we must determine the efficiency with which our observations find binary stars, given our discovery criterion. We use Monte Carlo simulations that mimic our observations for a population composed entirely of binaries. We create many (1000 – 10,000) artificial datasets with the same number of measurements per star, spacing of the measurements in time, and measurement uncertainties as the real data. Each "observed" giant is the primary of a binary system whose period, $P$, eccentricity, $e$, and secondary mass, $M_2$, are drawn from various distributions which we will describe later. The inclination of the orbit to the line of sight, the orientation of the major axis in the orbital plane, and the time of the first observation are chosen at random. The fraction of binaries in the simulations with $\chi^2$ probabilities below 0.001 is our efficiency. Unfortunately, this efficiency will depend on the properties of the input binary population and so we try several different possibilities to judge the impact of these assumptions on our final binary frequencies.

## 3.1.  Details of the Efficiency Simulations

The details of our simulations are similar to those of Pryor *et al.* (1988) and Sec. 2.1 of Hut *et al.* (1992). In particular, if the giant in a binary chosen for the simulation has a radius larger than its Roche lobe radius, as given by Eq. 2 in Eggleton (1983), then that binary is discarded and a new one is selected. Once mass transfer begins in a system, it is expected to be quickly removed from the magnitude-limited radial velocity sample by truncation of the evolution of the giant or by a common-envelope stage. For non-circular orbits, we evaluate the Roche radius at pericenter. This calculation is not exactly correct, but should be a reasonable approximation to the complex true situation.

Some of the stars in our sample are on the asymptotic giant branch (AGB) and for these it is more appropriate to compare the Roche radius to the largest radius these stars attained on the first ascent of the giant branch. On average, the available photometry is not precise enough to unambiguously determine whether a given star is on the first-ascent red giant branch (RGB) or the AGB. Distinguishing between the two is further complicated by the wide giant branch of Draco (Carney & Seitzer 1986) and the possibly wide giant branch of UMi (Olszewski & Aaronson 1985). We therefore took into account the existence of AGB stars in the following way. Buonanno *et al.* (1994) determined that the ratio of AGB to RGB stars above the horizontal branch is 0.24 in the globular cluster M3, which implies that about 19% of the stars above the horizontal branch are on the AGB. Some of our simulations suggested that this observed value might be reduced to $0.95\times$ the true value because of mass transfer in binary systems. Thus, each star in our simulations has a 20% chance of being classed as an AGB star and assigned a radius equal to that at the tip of the first-ascent RGB.

Mass loss is, in principle, a further complication for the AGB (and RGB) stars. By the time Population II stars reach the horizontal branch, they have lost 0.1–0.2 M$_\odot$, probably through strong stellar winds near the tip of the giant branch. If the mass loss occurs slowly compared to the orbital period and carries off no angular momentum (the first of these assumptions is likely, the second is more debatable), then the orbit expands while preserving its shape. One effect of this is to allow AGB stars to avoid mass transfer in initially more tightly bound binaries than would otherwise be the case. Treating this orbital evolution correctly would require integrating the mass loss along the giant branch and comparing the changing orbital separation to the evolution of the radius of the giant. The mass loss history is not well known, so we simply assume that the loss is concentrated near the time when the giant is at the RGB tip and is already at nearly its maximum size. We thus compare the tip radius to the Roche radius of the orbit before mass loss. This eliminates the maximum number of short-period orbits and thus yields a minimum efficiency. Reducing the radius assigned to the AGB stars from 0.4 AU to 0.3 AU increases the discovery efficiency by only a factor of 1.05, so the effect of this approximate treatment on our results is small.

Mass loss also affects the detectability of binaries containing an AGB star through its effect on the velocity amplitude of the star. Given the assumptions of the previous paragraph, it is straightforward to evaluate the change in the velocity amplitude using the two adiabatic invariants for a spherically symmetric potential, the radial action and the angular momentum (Lynden-Bell 1973). If $M_1$ and $K_1$ are the original primary mass and velocity amplitude and $M_1'$ and $K_1'$ are the values after mass loss, then

$$\frac{K_1'}{K_1} = \left(\frac{M_1 + M_2}{M_1' + M_2}\right)\left(\frac{M_1'}{M_1}\right)^{1/2}. \tag{1}$$

The decrease in the amplitude caused by the increasing orbital separation is offset by the increase caused by the AGB star moving further from the center of mass. The amplitude can either increase or decrease, depending on the initial mass ratio, but changes by less than 5% for the ranges of binary parameters and amount of mass loss that are of interest. We tested incorporating this complication into our efficiency simulations and found that it changed the efficiency by a factor of 1.004 or less. Thus we chose to ignore this effect; unless explicitly noted otherwise, in all of the simulations that we report here we set the mass of the giant, $M_1$, to 0.8 M$_\odot$.

We derived radii for each star in our sample in the following manner. First, we adopted the isochrone from Bergbusch & VandenBerg (1992) with an abundance [Fe/H]=$-2.03$ and an age of 16 Gyr. This abundance closely matches the measured metal abundance of Draco and Ursa Minor (Lehnert $et$ $al.$ 1992; Suntzeff $et$ $al.$ 1984). We then calculated radii from the luminosities and effective temperatures given by the isochrone and derived an empirical relation between these radii and $M_V$. Apparent $V$ magnitudes for each of our 118 stars were taken from the photometry sources discussed in Sec. 2. As the adopted Bergbusch & VandenBerg (1992) isochrone implies $M_V$(HB) = $+0.50$ (Dorman 1992), for consistency we use apparent distance moduli, $(m - M)_V$, of 19.3 for UMi and 19.6 for Draco. These moduli are similar to the previously-derived moduli of 19.1 for UMi (Cudworth $et$ $al.$ 1986) and 19.5 for Draco (Stetson 1979). Finally, we inferred a radius for each star in our sample from its absolute visual magnitude. The efficiencies from our simulations change by a factor of 1.06 if we change the distance moduli by 0.1.

As a check on these radii, we calculated a radius for each Draco/UMi giant measured in the infrared by Aaronson & Mould (1985) using their observed luminosities (adjusted for our slightly different distance moduli) and effective temperatures. The $M_V$-radius relation from the theoretical models falls within the scatter of these observational values, giving us confidence in the model-based radii.

## 3.2. What Binaries Can the Draco/UMi Data Find?

We begin our discussion of the discovery efficiency for our Dra/UMi sample by surveying its dependence on period, ellipticity, and secondary mass (actually, on the mass ratio, $q = M_1/M_2$). Figure 3 plots efficiency *vs.* orbital period for the $\chi^2$ probability less than 0.001 discovery criterion. These are average efficiencies calculated in four bins per decade of period (equally spaced in $\log(P)$). Panel (a) shows the results for simulations that used only binaries with circular orbits and panel (b) shows them for simulations that used a population of binaries with a thermalized distribution of eccentricities, $f(e) = 2e$. The four pairs of dashed and solid curves in each panel are the discovery efficiencies in each of four equal intervals of the logarithm of the mass ratio in the range $0.33 > \log(q) > -1.0$. These curves are labeled by the average value of $q$ for the bin. The solid curves ignore the effect of mass transfer in binaries and would be the discovery efficiency for a sample of stars at the main-sequence turnoff. The dashed curves show the bias introduced by the elimination of close binaries from our magnitude-limited sample by mass transfer. The plotted efficiencies were calculated from 1000 simulated samples in each of the 24 period–mass ratio bins. Within each bin, binaries were distributed uniformly in $\log(P)$ and $\log(q)$.

Figure 3(a) shows that our Draco/UMi sample is best at discovering binaries with circular orbits that have periods of 1–2 yrs and has reasonable sensitivity to systems with periods between about 0.4 and 10 yrs. Figure 3(b) shows that the corresponding numbers for a thermalized eccentricity distribution are 4 yrs and 1.0–20 yrs. In either case we are very unlikely to detect systems with mass ratios below 0.2. The few preliminary periods given in Sec. 2 fall within the above ranges. The decreasing velocity amplitudes and slower rates of change of long-period orbits are what cause the efficiencies to fall with increasing period. The tail of sensitivity to orbits with periods beyond 10 yrs is due to the 36% of our sample with both MMT and Hydra observations, which span a median baseline of 8.4 yrs and a maximum baseline of 11.9 yrs. The dashed lines in Fig. 3 show that the decrease in efficiency at short periods is due to the lack of close binaries in our sample caused by the inability of the giants to fit within them. Naturally, this effect is more important for elliptical orbits, in which the stars approach more closely for a given period. The efficiency is also reduced for elliptical orbits because the stellar velocities change slowly for a large fraction of the period around the time of apocenter.

Figure 3 looks similar to equivalent plots for observations of bright globular cluster giants, *e.g.*, Fig. 1 of Hut *et al.* (1992). The efficiencies for the Draco/UMi sample are generally about a factor of two lower than those for the best-observed samples of cluster giants because of fewer observations per star and larger velocity uncertainties (the dSph giants are typically ≥3 magnitudes fainter than the globular cluster giants). In both the globular cluster and dwarf spheroidal samples, the peak discovery efficiency is $2\times - 3\times$ lower for a population of binaries with a thermalized distribution of eccentricities than for a population with circular orbits for the reasons described above.

### 3.3. Average Discovery Efficiencies for a Restricted Binary Population

Until the distribution of binary properties is known, there is no one correct answer to the question of how to average the binary detection efficiency over those properties. The average efficiency and the resulting true binary frequency will clearly depend on how the averaging is done. All that we can do is to use a variety of approaches, thus illustrating the range of possible results, and to state clearly what we have done. We will use two general approaches: 1) averaging the curves of Fig. 3(a) and (b) over various ranges of period and secondary mass, which is discussed in this section, and 2) determining average efficiencies by performing separate simulations that use the distribution of binary properties found for nearby solar-type stars by Duquennoy & Mayor (1991; DM hereafter), which is discussed in Sec. 3.4.

The first two rows of Table 2 give the result of averaging the curves for the three bins with the largest mass ratios in Fig. 3 (the three highest dashed curves) over the interval 0.4–8 yrs for circular orbits and the interval 1.0–20 years for orbits with thermalized eccentricities, respectively. The first column of the table describes the averaging and the second lists the average binary discovery efficiency. The third and fourth columns give the resulting true binary frequencies, with their 95% confidence intervals, for discovery fractions of 0.051 and 0.034, respectively. The confidence intervals reflect only the uncertainties in the discovery fractions. The uncertainty due to the averaging can be judged by comparing the frequencies from different rows of Table 2.

The meaning of the entry in column 3 of the first row of Table 2 is that 0.17 is our best estimate of the frequency with which stars in Draco and UMi near the main-sequence turnoff (before mass transfer has acted on the giant branch) are the primaries of a binary system with a period between 0.4 and 8.0 yrs and a mass ratio between 0.22 and 2.1. This estimate assumes that the orbits are nearly circular. The entry in the next row, 0.47, gives our estimate for the frequency of binaries with periods in the range 1–20 yrs and mass ratios between 0.22 and 2.1 on the assumption that the orbits have a thermalized distribution of eccentricities.

The binary frequencies derived as described in the previous paragraph clearly depend on the range of periods employed in the average. They would decrease if the range were narrowed and increase if it were widened. One way to reduce this sensitivity is to make the assumption that there are, over the range of periods to which our data are sensitive, equal numbers of binaries in equal intervals of $\log(P)$. Since our range of sensitivity is at most two decades and studies of field main-sequence stars, such as DM, show that the periods of binary stars extend over 11 decades, this assumption is probably not too bad. With this assumption, integrating under the average of the three curves in Fig. 3 for the larger mass ratio bins yields an "equivalent coverage", $\Delta \log(P)_{100}$, which is the size of the interval in $\log(P)$ that would detect the same number of binaries if the efficiency were 100%.

Dividing the discovery fraction by $\Delta \log(P)_{100}$ yields the binary frequency per decade of period. Rows three and four of Table 2 show the result of doing this for circular and thermalized orbits, respectively. These frequencies are our estimate of the fraction of stars near the turn-off that are primaries of systems with mass ratios between 0.22 and 2.1 and with periods in an arbitrary decade of period. The frequencies are similar to those in the first two rows if one remembers that the latter are for 1.3 decades of period.

The average efficiencies and the binary frequencies described above were calculated in a similar fashion to those given in Sec. 2.1.3 of Hut *et al.* (1992) for a sample of globular cluster giants. We will compare the binary frequencies in various environments in Sec. 3.6. Here we instead consider whether this scheme places too much weight on the efficiencies for mass ratios near 1.0. DM found that such binaries are quite rare among nearby solar-type stars (see their Fig. 10). Their most likely $q$ was about 0.23. The low stellar densities in Draco and UMi suggest that, unlike the situation for globular clusters, their binary populations will have been little altered by stellar encounters since a time shortly after the stars formed. Thus the distribution of secondary masses might be similar to that for the Pop I field and a better estimate of the true binary fraction might result from an average efficiency weighted to that for lower mass ratios.

Counting systems in Fig. 10 of DM shows that there are roughly equal numbers of binaries in the intervals $-0.66 < \log(q) < -0.33$ and $-0.33 < \log(q) < 0.00$ and very few at larger $q$. There are significant numbers of systems with smaller values of $q$, but, for the moment, we restrict ourselves to the ranges of binary properties to which our data are sensitive. We thus averaged the middle two efficiency curves in Fig. 3. The next four rows in Table 2 show the results of doing this using parameters that are otherwise the same as for the first four rows. The efficiencies are about 30% lower than when $q$'s greater than 1.0 are included and the binary frequencies correspondingly higher.

### 3.4.  Average Efficiencies for a DM Binary Population

Finally, we calculated average efficiencies by adopting the distributions of period, eccentricity, and mass ratio found by DM. Thus, in these Monte Carlo simulations, the binary periods were selected from the distribution

$$f(\log(P/yr)) \propto \exp\left(\frac{-(\log(P/yr) - 2.24)^2}{2(2.3)^2}\right) \qquad (2)$$

for $-3.56 < \log(P/yr) < 7.44$ and the mass ratios from

$$f(q) \propto \exp\left(\frac{-(q - 0.23)^2}{2(0.42)^2}\right) \qquad (3)$$

for $q > 0$.

DM found that the distribution of eccentricities varied with period, with the orbits being circular at the shortest periods (due to tidal circularization) and the fraction of eccentric orbits increasing as the period increased. For $P > 2.7$ yrs ($1000^d$), DM argued that the distribution was tending towards $f(e) = 2e$. However, the histogram in their Fig. 6b contains fewer large eccentricities than would be predicted by that distribution. Thus, for this period range, we chose eccentricities from the distribution $f(e) = 1.5e^{1/2}$, which is a reasonable match to the observed distribution corrected for selection biases. For shorter periods, we chose eccentricities from a distribution that reproduced the histogram in DM Fig. 6a. This distribution is sharply peaked around $e \simeq 0.3$ and is zero for $e > 0.75$. Because of the elimination of short-period binaries from our sample by mass transfer, the circular-orbit regime found by DM is irrelevant for our simulations.

Tidal circularization can also act on binaries containing a giant with separations somewhat larger than those which produce mass transfer. This is probably the reason why Dra C1 appears to have at least an approximately circular orbit. We neglect this complication in our simulations since it probably affects only a small range of the periods that are not already eliminated by mass transfer.

The next-to-last row of Table 2 gives the average efficiency and the corresponding binary frequencies derived from simulations of 10,000 Dra/UMi samples using the DM distribution of binaries. Because our data are sensitive to only a small fraction of the total period range of Eq. 2 and because the DM distribution of mass ratios peaks at small values, the average efficiency is small, 0.0247. This produces binary frequencies that are larger than 1.0. We emphasize that these results assume that the population of binaries in Draco and UMi has the same properties as that studied by DM, which may or may not be true. The derived binary frequency is for a much wider range of period than those given above it in Table 2 and entails a large extrapolation from the relatively narrow period range to which the data are sensitive.

A calculated binary frequency as large as 2.0 probably does suggest that one or more of the DM distributions does not apply to the binary stars in Draco and UMi. However, we note that a binary frequency modestly greater than 1.0 is possible. We define the binary frequency as the fraction of the primaries that are members of a binary system with periods in a certain range. The primary of a triple system has two periods and that of a quadruple system has three. DM found that 9 of the 71 binary systems that they identified in their complete sample contained three or more stars.

Finally, we have performed Monte Carlo simulations using a more limited period range and the DM distributions with the goal of deriving a binary frequency more directly comparable to those derived earlier and thus exploring the impact on the efficiency of adopting the DM mass ratio and eccentricity distributions. In our original DM simulation, 95% of the binaries with $\chi^2$ probabilities less than 0.001 had periods between 0.5 yrs and 30 yrs.[3] The final row of Table 2 gives the efficiency and frequencies that result when the binary periods are drawn from Eq. 2, but are required to be between 0.4 yrs and 40 yrs. We compare the various binary frequencies in Table 2 in the following discussion.

### 3.5. A Cautionary Discussion of Binary Frequencies

We emphasize that the wide range of binary frequencies in column 3 or column 4 of Table 2 are derived from the same data. The range of values comes partly from the different period and secondary mass ranges to which the frequencies refer and partly to the different assumptions made in the averaging. We can approximately remove the period dependence by dividing those frequencies that are not already expressed per decade of period by the number of decades to which they refer. The frequency per decade for a discovery fraction of 0.051 is then 0.1–0.2 for circular orbits, 0.3–0.5 for thermalized eccentricities, and 0.2–0.3 for the DM eccentricity distribution. The frequency increases as the mean eccentricity of the binary population used to calculate the efficiency increases, reflecting the difficulty of detecting binaries with very eccentric orbits with just a few radial velocity observations.

---

[3]The upper limit of this period range has been corrected to remove the effect of "detected" long-period systems that actually have constant velocities but which satisfied our discovery criterion by chance. Such systems are about 4% of the binaries "detected" in this simulation (the ratio of 0.001 to 0.0247). There is no equivalent bias on the short-period side of the range because those systems are eliminated by mass transfer.

Our ignorance of the eccentricity distribution for the binaries in Draco and UMi makes the largest contribution to the uncertainty in the binary discovery efficiency. The uncertainty in the efficiency makes a contribution to the uncertainty in the binary frequency that is comparable to that from small number statistics in the discovery fraction. A more secure binary frequency for the dSph galaxies will require both more velocities for the binary candidates already discovered and increasing the number of velocities and their extent in time for the sample as a whole.

The binary frequency of 0.2–0.3 per decade of period derived using the DM distributions of $P$, $q$, and $e$ should probably be considered as our present best estimate of the number of binaries among stars near the turnoff in Draco and UMi. This estimate obviously only applies to binaries with periods not too distant from 1 year. While this fraction may seem high, we note that many of these systems are expected to have very low secondary masses. Such systems will be very difficult to detect and will have negligible effect on the measured kinematics of Draco and UMi. We will return to the latter point in Sec. 4.

### 3.6. Comparing Binary Frequencies from Different Environments

We have derived the first binary frequency for a dSph galaxy. How does the frequency for Draco and UMi compare with those for other stellar environments? The simplest comparison is probably with the DM survey of nearby solar-type stars. After correcting for incompleteness, DM had 100 periods for their complete sample of 164 primaries. This is a binary frequency, as we have defined it in this paper, of 0.61. This should be compared to our binary frequency which spans the same range of period and secondary mass and was derived assuming the DM $P$, $e$, and $q$ distributions. If UMi N33 and Dra 249 are binaries then our value is 2.1, with a 95% confidence interval of (0.97, 3.9). Excluding UMi N33 and Dra 249 yields 1.4 with the 95% confidence interval (0.38, 3.4). Thus Draco and UMi appear to have more binaries than the DM sample, with our best estimate being that the dSphs have about 3× more. Excluding equal binary frequencies with high statistical confidence depends on the status of UMi N33 and Dra 249, however.

The above comparison assumes that the DM period distribution is valid for the Draco and UMi binaries over a much wider period range than that to which our data are sensitive.

A more conservative procedure is to compare the binary frequencies for binaries to which both surveys are sensitive. DM found 19 systems with periods between 0.4 yr and 40 yr and any secondary mass. This should be increased by about one for incompleteness, yielding a binary fraction of 0.12 for that period range. Our value is 0.60, with a 95% confidence interval of (0.28, 1.1), or 0.40, with an interval of (0.11, 0.99). Our much larger binary frequencies suggest that the period, eccentricity, or secondary mass distributions of the Draco and UMi binaries are significantly different than those of DM. Given the low density of the stellar environment in both the solar neighborhood and the dwarf spheroidals, we find this result surprising.

Hut *et al.* (1992) report a binary frequency for a sample of 393 globular cluster giants of 0.05 for circular orbits and 0.12 for a thermalized distribution of eccentricities. These frequencies apply to the two decades of period between 0.2 yrs and 20 yrs and for mass ratios between 0.22 and 2.2. They are thus, except for a slightly wider period range, comparable to the binary frequencies in the first two rows of Table 2. There are some differences in the details of the way the efficiencies were calculated in the two studies, but this should not significantly affect the comparison. Adjusting for the different period ranges, our binary frequency for Draco and UMi is about $5\times$ larger than that found for globular clusters. This difference suggests that whatever is the cause of the large binary frequencies that we have found in Draco and UMi, it not simply a difference between Population I and Population II.

## 4. Effects of Binaries on the Velocity Dispersion in Published dSph Samples

We are now in the position to use the binary fraction derived above, or to use the individual velocities in Draco and Ursa Minor, to assess the importance of binaries in all of the dwarf spheroidals with published velocity dispersions. In this section we will make three distinct types of tests. In Sec. 4.1, we will calculate dispersions for subsamples of stars with different numbers of observations and with differing cuts on the uncertainties in the velocities. Those data with larger numbers of epochs and smaller velocity uncertainties make it easier to recognize and eliminate binaries. Therefore, by calculating dispersions based on subsamples of increasingly better-quality data, we explore the sensitivity of the dispersions to the effects of binaries. In Sec. 4.2, we will use the velocities in our Draco and UMi samples to calculate a mean velocity for each star and then find the distribution of the deviations from these means. The distribution of velocity deviations contains information about both the known and the undiscovered binaries in Draco and UMi. To the extent that Draco and UMi have representative binary populations, we can apply those deviations to simulate single-epoch observations in any other dwarf spheroidal. The resulting velocity dispersion measures the effects of the binary population. In Sec. 4.2, we also perform the same procedure but draw the velocity deviations from the DM distribution of field binaries, properly normalized to our data.

Although much work has gone into deriving dwarf spheroidal velocity dispersions, the published data on multi-epoch observations is still small and the time differences between different epochs is generally not large. Our UMi/Dra data is the best sample, with a variety of epoch spacings and the longest baseline. Table 3 gives the number of stars with maximum epoch differences larger than various limits for the stars with multiple observations in UMi and Draco. Sculptor and Sextans now have observations taken at only two or three different epochs. The Sculptor data were taken in 1985 (Armandroff & Da Costa 1986) and 1990 or 1991 (Queloz *et al.* 1995). There are 15 stars in common between these two studies. The Sextans data were obtained in 1991 (Suntzeff *et al.* 1993) and by Hargreaves *et al.* (1994a) in 1991 (at about the same time as the Suntzeff *et al.* study) and 1992. Twelve stars are in common.

The above multi-epoch studies are less affected by binaries than the single-epoch studies. In Sec. 4.2, we will model the effect of binaries on the single-epoch data in Sculptor (Queloz *et al.* did not combine the data), Leo II (Vogt *et al.* 1995), Carina (Mateo *et al.* 1993), and Fornax (Mateo *et al.* 1991).

## 4.1. Subsamples

We can use our Draco and Ursa Minor data to select subsamples of stars from which to derive velocity dispersions. These subsamples, which have different velocity uncertainty limits and different numbers of repeat observations, give us differing abilities to weed out binaries and differing susceptibility to the effects of binaries. Thus, the changes in velocity dispersion between these subsamples is a measure of the effects of binaries on the velocity dispersion.

Table 4 presents velocity dispersions for subsamples of the Draco and UMi velocities. The second column of each row gives the minimum number of epochs necessary for a star to be in that sample. Four pairs of columns follow, each giving the number of stars and the dispersion with its uncertainty for a different subset of the stars with that number of observations. The first and second subsets are all of the stars and all of the stars except for known velocity variables. The third and fourth subsets contain only those stars for which every velocity is more accurate than 3.5 km s$^{-1}$, with the fourth also having the known binaries eliminated.

The first set of four rows in Table 4 gives the velocity dispersions for Draco when the three extreme-velocity stars (see Paper I) are excluded from the sample. The second set of four rows gives the dispersions when these stars are included. A third set of four rows gives the results for UMi, excluding the one extreme velocity star (this star had only one measured velocity with an uncertainty of 3.6 km s$^{-1}$). What is striking about Table 4 is that all of the dispersions are virtually the same to within their uncertainties. The most extreme difference for UMi, between the dispersion for the 42 stars observed once or more and with uncertainties less than 3.5 km s$^{-1}$ and the dispersion for the 18 stars observed twice or more with uncertainties less than 3.5 km s$^{-1}$ and with binaries removed, is 1.4$\times$ the combined uncertainties. For Draco without the extreme velocity stars, the minimum and maximum values of the dispersion are 8.29$\pm$0.84 and 10.22$\pm$1.87 km s$^{-1}$; these are separated by 0.9$\times$ their combined uncertainties. The largest and smallest dispersions for the Draco sample with extreme velocity stars added are 1.4$\sigma$ apart. The ability to remove known binaries and the additional certainty gained by multiple epochs make little difference in these measured dispersions.

Another type of sample that can be created from the complete datasets published in Paper I is to draw one velocity for each star at random. This Monte Carlo simulation tests whether observing at a random epoch will skew the resulting UMi and Draco velocity dispersions. Ten thousand realizations of single-observation samples in both Draco and UMi were created. Each sample contains one velocity for each star. A star with only one measured velocity has that same velocity used each time, while a star with $N$ velocities has each velocity used at random $1/N^{th}$ of the time. We calculated the maximum-likelihood velocity dispersion for each sample using the uncertainty for each individual velocity (Paper I). Table 5 gives the results of this test, expressed as the number of trials falling in intervals around the published (Paper I) dispersion defined by the uncertainty in that dispersion. The vast majority of all samples have dispersions within 1$\sigma$ of the full-sample dispersion. We then performed the same test removing the one extreme-velocity star in UMi and the three extreme-velocity stars in Draco. The results from these tests are also found in Table 5 and, again, almost all of the samples fall within 1$\sigma$ of the full sample dispersion. Clearly, the errors in the measured velocity dispersions in Draco and UMi due to binaries are smaller than the sampling uncertainty in the dispersions.

From these two experiments we conclude that binary stars, whether detected or not, or removed or not, make little difference in the derived properties of the UMi and Draco dwarf spheroidal galaxies. We will show below that they also make little difference in the derived velocity dispersions of the other dwarf spheroidals.

## 4.2. The UMi/Dra Stars with Multiple Velocities and Binaries in Other Dwarf Spheroidals

Some dwarf spheroidals have smaller velocity dispersions than those of Draco and Ursa Minor. Sculptor, Sextans, Leo II, and Carina have published dispersions in the range 6–7 km s$^{-1}$. These particular galaxies are more likely to have their velocity dispersion skewed to higher values due to the presence of binaries. Have they been significantly affected? We can use both the velocities of stars in Draco and UMi and the derived binary frequency coupled with the properties of solar neighborhood binaries to ascertain the effects of binaries on any measured velocity dispersion in a dwarf spheroidal.

We begin with tests that are not based on a model for the binary population. Instead, these tests employ the scatter in the multi-epoch Draco/UMi observations. This scatter contains the effects of the obvious binaries discussed in Sec. 2 above and the effects of any undiscovered binaries as well. The tests incorporate only a single assumption: that the binary populations in Draco and UMi are representative of the binaries in the other dwarf spheroidals.

The deviations employed in the simulations were derived from a set of 61 stars with two or more velocities and with all velocities more accurate than 3.5 km s$^{-1}$ (19 stars have 2 velocities, 12 have 3, 9 have 4, 10 have 5, 8 have 6, and 3 have 7 velocities). These accurate velocities all have approximately the same uncertainties and so we can calculate unweighted mean velocities and the deviations from these means. Using weighted means with a wider range of velocity uncertainties would introduce spuriously small deviations in those cases when pairs of velocities with very unequal weights were averaged. The value of 3.5 km s$^{-1}$ was chosen because it is 2× the median single-observation uncertainty in the MMT dataset, which is the most accurate of the data discussed in Paper I. The empirical deviations were increased by a factor of $N/(N-1)$, where $N$ is the number of observations of a star, to account for the small sample bias. We have also derived a separate set of velocity deviations from the 54 stars excluding the C stars in order to simulate samples in which C stars were specifically excluded.

We first use the deviations to derive the velocity dispersions contributed by binaries alone in samples of 20, 30, 40, 50, and 60 stars. These dispersions are one way to quantify the magnitude of the effects of binaries. Ten thousand trials at each sample size were carried out. For each sample we randomly picked the same number of observed stars. An individual star could be included more than once. Then, for each star we picked an individual deviation at random. Table 6 contains five columns giving the results for the five different sample sizes. The first two rows give the median velocity dispersion calculated using the maximum likelihood method and the 95% upper bound of the trials, respectively.

The second pair of rows gives the same quantities for the biweight dispersion with the average velocity uncertainty subtracted in quadrature.

The dispersions contributed by binaries alone in Table 6 are all between 1.2 and 1.8 km s$^{-1}$, with 95% upper bounds of 2.0–4.8 km s$^{-1}$. Unless the authors of the published velocity dispersions of Carina (6.8 km s$^{-1}$, Mateo *et al.* 1993), Leo II (6.7 km s$^{-1}$, Vogt *et al.* 1995), Sculptor (6.3 km s$^{-1}$, Armandroff & Da Costa 1986; 6.2 km s$^{-1}$, Queloz *et al.* 1995), and Sextans (6.7 km s$^{-1}$, Suntzeff *et al.* 1993; see their Table 8, line 1) were all exceedingly unlucky, removing the effects of binaries would lower the typical velocity dispersion only from 6.7 to 6.5 km s$^{-1}$. Again, binaries are relatively unimportant.

A similar test is to duplicate the exact sample sizes and velocity dispersions of the above studies by drawing velocities from a Gaussian with a width equal to the published velocity dispersion. We then add deviations chosen at random from the Draco/UMi sample to each velocity. Finally, Gaussian noise is added to increase the measurement uncertainties from those in the higher-quality Draco/UMi sample to those in the published study being investigated. Both maximum likelihood and biweight dispersions were measured for ten-thousand samples drawn in this fashion.

There are four columns in Table 7, which gives the results of these simulations. The first column lists the name of the galaxy, and, under it, the number of stars used in deriving the published dispersion and the mean velocity uncertainty per star. Some galaxies are listed more than once with the differences in the samples explained in the table notes. The remaining three columns give results for three different simulations. The first line in each of those columns shows the input velocity dispersion. The second line provides the median and 95% upper bound to the dispersions derived using maximum likelihood, while the third line gives the same quantities derived using the biweight with the mean uncertainty subtracted in quadrature. Simulation 1 uses the published velocity dispersion as input. Simulation 2 uses the velocity dispersion whose 95% upper bound maximum-likelihood dispersion equals the published dispersion. Simulation 3 is the same as Simulation 1, except that the C stars have all been excised from the table of input deviations; most modern observational studies deliberately exclude C stars.

Simulations 1 and 3 give median values that are very close to the input velocity dispersion, with worst-case differences of 0.2–0.3 km s$^{-1}$. These differences are small compared to the statistical uncertainties of the velocity dispersions for these samples, which range from 1.1 to 2.3 km s$^{-1}$. Figure 4 shows the histogram of the derived maximum likelihood dispersions for Simulation 1 of the Carina galaxy, thus displaying how the resulting dispersions are distributed. Note the concentration of the values around the input dispersion. Simulation 2, which can be considered the worst-case lowest velocity dispersion for each galaxy, and surely cannot be relevant for more than one of the galaxies considered here, changes the velocity dispersion by 30–50%.

The above simulation is not strictly correct when applied to the Fornax and Carina dwarf spheroidals. Both galaxies are known to have a substantial intermediate-age population as well as the classical old population (Aaronson & Mould 1980, Mould $et$ $al.$ 1982, Frogel $et$ $al.$ 1982, Mould & Aaronson 1983, Mighell 1990, Mighell & Butcher 1992, Smecker-Hane $et$ $al.$ 1994, Beauchamp $et$ $al.$ 1995). A recent color-magnitude diagram of Fornax (Beauchamp $et$ $al.$ 1995) confirms that some of the stars in Fornax are as young as 2 Gyr and shows that there is a substantial population between 2 and 5 Gyr. The most massive primary stars that are likely present in the radial velocity work of Mateo $et$ $al.$ (1991) would have velocity variations larger than those in UMi/Dra by a factor that averages 1.18 for a given period. This factor is not large enough to affect our conclusions significantly, especially when one considers that this massive population is not dominant.

Carina, with a measured velocity dispersion of 6.8 km s$^{-1}$ and a turnoff mass of about 1.0 M$_\odot$ for its 6.2 Gyr population ($e.g.,$ Smecker-Hane $et$ $al.$ 1994), will have its binary-star velocity amplitudes increased by a factor that averages 1.08. The more massive stars present in the Carina and Fornax velocity samples do not affect our simulations in an appreciable way. The measured velocity dispersions are not affected by binaries in these dwarf spheroidals unless the binary populations are radically different from those in UMi/Draco.

Our last experiment uses the binary frequency that we derived in Sec. 3.4 using the DM distribution of eccentricities, secondary masses, and periods. As in the previous simulations, samples identical to those in the published single-epoch studies were drawn from the model binary population. These stars were then "observed" at a random time and the resulting velocities added to a set of velocities drawn from a Gaussian. Because the binary frequency derived from the entire DM period range was larger than 1.0, the simulations used a period range for which the frequency would be 1.0 if the binary population had the DM period distribution. Thus, periods were chosen from Eq. 2 between the limits of 0.3 and 550 yrs. As in Sec. 3.4, systems that would have undergone mass transfer were eliminated from the sample.

Table 8 contains the results of these simulations and is laid out in the same way as Table 7. There are three columns: the first again identifies the galaxy or sample of stars, with the second line giving sample size and mean velocity uncertainty per star. Columns two and three are the two simulations. The first uses the published dispersion as the input dispersion, while the second uses the velocity dispersion which gives the published dispersion as the 95% upper bound to the maximum-likelihood calculation. The second line of columns two and three gives the median and 95% upper bound dispersions calculated using maximum likelihood, while the third line does the same for those calculated using the biweight with the mean uncertainty subtracted in quadrature. The results in Table 8 look remarkably like the results in Table 7. On average, binaries make little difference. At the 95% level, binaries change the derived dispersions of the galaxies with dispersions of 6–7 km s$^{-1}$ by 33–53%. However, the most likely changes are much smaller, a few percent. The odds that the dispersions of Carina, Leo II, Sculptor, and Sextans have all been inflated to the 95% levels are $6 \times 10^{-6}$.

### 4.3. Comparison with Others' Dispersion Simulation Results

Our simulations show that the effects of observed and modelled binaries are quite small. Are these results in accord with others' simulations? Aaronson & Olszewski (1987) show the results of two simulations carried out for them by R. Mathieu. Ten stars were each observed one year apart, and all stars with a velocity variation of greater than 4 km s$^{-1}$ were excluded from the calculation of the velocity dispersion. A binary frequency of 50% (50% of all visible giants were assumed to be binaries) was used; the period distribution, secondary mass distribution and eccentricity distribution are described in Mathieu (1985). Two simulations were made: the first had an input velocity dispersion of 1.5 km s$^{-1}$, while the second input dispersion was 10 km s$^{-1}$. We estimate from Fig. 2 of Aaronson & Olszewski (1987) that the median velocity dispersion from the first simulation is 1.5 km s$^{-1}$, with 95% lower and upper bounds of (0.3,5.0); the same parameters for the second simulation are 9.3 km s$^{-1}$, and 95% bounds of (5.5,13.7). We used our binary sample derived from the comparison with DM to mimic this simulation. For an input dispersion of 1.0 km s$^{-1}$, these simulations yielded a mean observed dispersion of 1.9 km s$^{-1}$, with a 95% upper limit of 3.4 km s$^{-1}$; for an input dispersion of 10 km s$^{-1}$, the results were a mean of 9.8 km s$^{-1}$, with a 95% upper bound of 14.3 km s$^{-1}$. Our new simulations thus agree very well with the simulations derived from Aaronson & Olszewski's (1987) early sample of velocities in UMi/Dra.

Tremaine (1987) notes that his Monte Carlo simulations based on Aaronson's (1987) UMi/Dra velocities give a dispersion due solely to binaries of 10$f$ km s$^{-1}$, where $f$ is the fraction of stars showing a $\geq$4 km s$^{-1}$ velocity difference between two observations separated by one year. The sample of stars described in OAH95 provides a good upper limit to the parameters of Tremaine's simulation. Nine stars of the 33 with multiple observations have velocity extrema (not simply 2 observations one year apart) of $\geq$5 km s$^{-1}$. This set of stars implies an upper limit to the dispersion of 2.7 km s$^{-1}$, which implies that binaries have had little effect on the measured dispersions of any of the dwarf spheroidals.

Mateo *et al.* (1993), Suntzeff *et al.* (1993), and Vogt *et al.* (1995) have published binary simulations similar to ours, showing the effects of binaries in both tabular and graphical form. These results have been extensively quoted (*i.e.*, Queloz *et al.* 1995; Hargreaves *et al.* 1994a,b). We were forced to compare quite carefully the results in this paper to the previously published results and, while doing so, we found some discrepancies. The Fortran routines which made the simulations referred to above (Mateo *et al.* 1993, Suntzeff *et al.* 1993, Vogt *et al.* 1995) were found to have a coding error in the way that the velocity of the of the primary of a binary was calculated. These velocities were overestimated by a value of $(M_1 + M_2)/M_2$, typically a factor of 3.0. The role of binaries was thus overestimated and the corrected simulations even more strongly support the conclusion of these papers that binaries are unlikely to have a significant effect on the measured dispersions.

The first 12 lines (Simulation 1) of Table 9 give the recalculated simulations of Mateo *et al.* (1993) and serve as a guide to the errors in the simulations in Suntzeff *et al.* (1993) and Vogt *et al.* (1995). These lines are for exactly the parameters in Table 10 of Mateo *et al.* (1993), specifically, for 0.8 M$_\odot$ primaries. Simulation 2 gives the results for a grid of binary frequency (0.1–0.7) and period distributions (minimum period of 0.5 years; maximum period of 100 or 1000 years). Simulation 3 shows the effects of a 1.0 M$_\odot$ primary and can be considered a Carina simulation. Finally, in Simulation 4 we show the velocity dispersion that gives the upper end of a 95% confidence interval to be equal to the Carina dispersion of 6.8 km s$^{-1}$, assuming binary frequencies of 0.2–0.9.

The corrected older Monte Carlo simulations (Mateo *et al.* 1993, Suntzeff *et al.* 1993, Vogt *et al.* 1995) are now in agreement with our current results. Specifically, if Carina were to have 30% binaries, then its velocity dispersion would be inflated from 4.7 km s$^{-1}$ to 6.8 km s$^{-1}$ or larger only 2.5% of the time. Even 90% binaries would inflate 1.6 km s$^{-1}$ (the velocity dispersion that gives M/L=2.0 for Carina) to 6.8 km s$^{-1}$ or larger only 2.5% of the time. Binaries make only a small difference in these derived dispersions. Finally, our results (Tables 7 and 8) for Carina are remarkably in accord with the results of Table 9 for 30% binaries. We remind the reader that our binary frequencies in Table 2 range from 0.17 to 0.60 for six detected binaries, and 0.12–0.4 for four detected binaries.

## 4.4.   Velocities in Sculptor as a Test Case

Queloz *et al.* (1995) have published second-epoch velocities in Sculptor, following the earlier work of Armandroff & Da Costa (1986). Queloz *et al.* observed 24 stars, 15 of which were in common to the first-epoch study. Of these 15 stars, two were found to have variable velocities and were discarded. After removing a velocity zeropoint difference of 3.5 km s$^{-1}$, Queloz *et al.* found that the standard deviation of the differences between the pairs of velocities for the remaining 13 stars was 5.1 km s$^{-1}$, completely in accord with the expected standard deviation of 5.0 km s$^{-1}$ from the measurement uncertainties added in quadrature.

If these two velocity variables had remained undetected, would the velocity dispersion vary by the small amounts that our simulations predict, or is there a major effect? Armandroff & Da Costa (1986) derived a velocity dispersion of $6.3^{+1.1}_{-1.3}$ km s$^{-1}$, including observations of these two stars. Queloz *et al.* (1995), eliminating these two stars, derive $6.2\pm1.1$ km s$^{-1}$. We derive $7.3\pm1.2$ km s$^{-1}$ if the Queloz *et al.* observations of these two stars are reinserted into the velocity dispersion calculation and $6.6\pm1.1$ km s$^{-1}$ if the (suitably zero-point shifted) Armandroff & Da Costa velocities for the two stars are used with the rest of the Queloz *et al.* sample. Comparing these last two dispersions isolates the effect of the two binaries on the dispersion the most clearly, since the sample of stars is identical in both cases.

The simulations summarized in Table 7 show that the $1\sigma$ scatter due to both binaries and the sampling uncertainty around an input dispersion of 6.2 km s$^{-1}$ is $^{+1.4}_{-1.1}$ km s$^{-1}$ for a sample of 21 stars and the binary velocity distribution of UMi and Dra. Similarly, the scatter is $^{+1.4}_{-1.3}$ km s$^{-1}$ for an input dispersion of 7.3 km s$^{-1}$ and a sample of 23 stars. These scatters are comparable to the difference between the dispersions of Armandroff & Da Costa (1986), 6.3 km s$^{-1}$, and Queloz *et al.* (1995), 7.3 km s$^{-1}$. However, this comparison is not completely fair as the two samples are not totally independent (15 stars are in common). Thus, we performed some additional simulations that used a single sample with a dispersion of either 6.2 km s$^{-1}$ or 7.3 km s$^{-1}$ and only changed the deviations added to the velocities by binary orbital motion from one trial to the next. The $1\sigma$ scatter around the input dispersion was $^{+0.7}_{-0.6}$ km s$^{-1}$ for both cases. Again, this is comparable the difference between 6.6 km s$^{-1}$ and 7.3 km s$^{-1}$.

We conclude that the results of Queloz *et al.* (1995), with or without the inclusion of binaries, are completely in accord with the results of Armandroff & Da Costa (1986). The differences between the different dispersions are also completely in accord with our simulations. The "large-velocity-amplitude" binaries in Sculptor do not significantly change the velocity dispersion, nor do "possible undetected velocity variables."

## 5. Conclusions

We have used our multi-epoch velocities in the Ursa Minor and Draco dwarf spheroidal galaxies to derive the binary-star content of dwarf spheroidals for the first time. The binary frequencies, derived from a discovery fraction of 0.051 (or 0.034), range from 0.17 to 0.68 (0.12 to 0.45) in the period range to which we are sensitive. These binary frequencies are formally 3–5× larger than those derived for the Population I field and for globular clusters. The largest sources of uncertainty in this binary frequency are the unknown distribution of eccentricities, the small number of variables discovered to date, and the uncertain status of a few of the binary candidates.

Despite the high binary frequency in Draco and UMi, we have presented simulations that show that the presence of binary stars in the current samples of giant stars with radial velocities in dwarf spheroidal galaxies causes errors in the derived velocity dispersion that are small compared to other sources of uncertainty, for instance, sampling statistics. These simulations use our knowledge of the mean velocities of stars in Draco and UMi and the deviations about those means in a model-independent way. We used the set of deviations to mimic all of the other single-epoch velocity samples in dwarf spheroidals. In the mean, there was little change in the derived dispersions whether or not binaries were included or removed from the samples. We also used our derived binary frequency for UMi/Dra with a model binary population to simulate the effects of binaries in other dwarf spheroidals. Again, the effects were minimal.

We thus conclude that, while further observations will improve our knowledge of the binary star frequency, the current samples of velocities yield velocity dispersions for the dwarf spheroidal galaxies that are not significantly affected by binary stars. Binaries are not responsible for the large observed mass-to-light ratios.

We thank Hugh Harris for sharing results in advance of publication and Mario Mateo for providing us with his simulation code and for discussions about that code. EO was partially supported by the NSF with grant AST 9223967; he thanks Peter Strittmatter for additional support. CP's research is supported by NSF grant AST 9020685. This paper was written while CP was on sabbatical at Steward Observatory and he thanks Peter Strittmatter for support during this time. EO would like to remember Marc Aaronson and Jerry Garcia with this work; they both played pivotal roles in aspects of this work.

Fig. 1.— A plot of the probability that a star with a constant velocity would exceed the observed $\chi^2$ value by chance *vs.* the absolute visual magnitude for the 118 stars with multiple observations in the combined Draco and UMi sample. We adopted apparent distance moduli of 19.6 for Draco and 19.3 for UMi. The tip of the giant branch should be at about $M_V = -2.7$. Arrows indicate stars with probabilities below 0.001.

Fig. 2.— Plots of radial velocity *vs.* Julian date for the six stars with $\chi^2$ probability less than 0.001. Note that the minor tick marks denote 5 km s$^{-1}$ intervals and that the major tick marks correspond to 20 km s$^{-1}$ intervals. The fitted sinusoid for Dra C1 has a period of 375.5 days, an amplitude of 5.77 km s$^{-1}$, and a mean velocity of –299.4 km s$^{-1}$; the fitted sinusoid for UMi N33 has a period of 216.7 days, an amplitude of 4.05 km s$^{-1}$, and a mean velocity of –246.1 km s$^{-1}$. These fits are only illustrative of the possible orbital periods of these two stars. Periods of less than 200 days are ruled out by the radii of Dra C1 and UMi N33.

Fig. 3.— Average binary discovery efficiencies as a function of orbital period for the Dra/UMi sample. The discovery criterion is a $\chi^2$ probability smaller than 0.001. The four pairs of solid and dashed curves are the average efficiency in four equal intervals of the logarithm of the mass ratio in the range $0.33 > \log(q) > -1.0$. These curves are labeled by the average value of $q$ for the bin. The solid curves ignore the effect of mass transfer in binaries while the dashed curves show the bias introduced by the elimination of close binaries from our magnitude-limited sample by mass transfer. (a) The average discovery efficiencies for binaries with circular orbits. (b) The average discovery efficiencies for binaries with a thermalized distribution of eccentricities.

Fig. 4.— Histogram of the derived maximum-likelihood velocity dispersions for Simulation 1 of the Carina galaxy (Table 7). Note that the $1\sigma$ error due to sampling statistics is 1.2 km s$^{-1}$.

Table 1: Velocities for Stars with $\chi^2$ Probabilities $< 0.001$.
This space intentionally left blank.

Table 2: Efficiencies and Binary Frequencies For the Combined Draco/UMi Sample.
This space intentionally left blank.

Table 3: Number of Stars with Various Epoch Differences.
This space intentionally left blank.

Table 4: Velocity Dispersions for Subsamples in UMi/Dra.
This space intentionally left blank.

Table 5: Velocity Dispersions from Resampling Single Observations of Draco or UMi Stars.
This space intentionally left blank.

Table 6: Median and 95-percentile Dispersions Due to Binaries for Subsamples of the UMi/Dra Velocities.
This space intentionally left blank.

Table 7: The Effect of Binaries on Single-Epoch Dispersions: Empirical Simulations.
This space intentionally left blank.

Table 8: The Effect of Binaries on Single-Epoch Dispersions: Model Simulations.
This space intentionally left blank.

Table 9: Simulations Using the Corrected Code of Mateo *et al.* (1993).
This space intentionally left blank.